\documentclass[aps,physrev,graphicx,amsmath,amssymb,reprint,superscriptaddress,nofootinbib]{revtex4-2} 
\usepackage{graphicx}
\usepackage{subfig}
\usepackage{mathtools}

\usepackage{dcolumn}
\newcolumntype{d}[1]{D{.}{\cdot}{#1} }
\usepackage{bm}

\usepackage{xcolor,soul}
\usepackage{physics}
\usepackage{tikz}
\usetikzlibrary{decorations.markings}

\usepackage{mathpazo} 

\usepackage{siunitx}

\draft 

\newcommand{\R}{\bm{R}}
\newcommand{\rt}{\bm{r}}
\newcommand{\Rt}{\tilde{\bm{R}}}
\newcommand{\rtt}{\tilde{\bm{r}}}

\begin{document}
\title{\textsc{Spindrift}: Learning quantum degeneracy from thermal purity in restricted path integral Monte Carlo}
\date{\today} 

\author{Jarvist Moore Frost}
\affiliation{Department of Chemistry, Imperial College London, Exhibition Road, London  SW7 2AZ, UK}
\affiliation{Department of Physics, Imperial College London, Exhibition Road, London  SW7 2AZ, UK}
\email[Electronic mail:]{jarvist.frost@imperial.ac.uk}


\begin{abstract}
Restricted path integral Monte Carlo (RPIMC) sidesteps the fluctuating Fermion sign problem by confining paths within nodal regions of a trial density matrix, thereby recovering polynomial scaling. 
However, this nodal surface must be provided from elsewhere; unless it is exact, it introduces a fixed-node energy error. 

Here we introduce \textsc{Spindrift}, a Variational Density Matrix approach that learns the many-body Fermionic density matrix from a regularised Bloch residual, evaluated on samples drawn by a restricted Worm algorithm. 
Motivated by the `purity' of quantum statistical mechanics at high temperature (where kinetic energy dominates), we train the density matrix along an imaginary-time (descending temperature) curriculum from an exact infinite-temperature heat-kernel starting point, learning the condensation of quantum correlations as temperature drops, through successive corrections to the previous reference. 
We parametrise our model with a permutation-equivariant continuous normalising flow to generate quasi-particle backflow trajectories, modulated by a symmetric Jastrow factor. 
This architecture guarantees exact Fermionic antisymmetry and spatial symmetry throughout training. 

Simulating $N=3$ interacting Fermions in a two-dimensional harmonic trap, we demonstrate stable curriculum training.
The learnt velocity field smoothly deforms the nodal structure away from the free-particle reference. 
Open-Worm G-sector trapping provides a natural diagnostic for nodal accuracy. 

\textsc{Spindrift} systematically lowers the restricted energy relative to the free-particle reference across all temperatures and successfully reproduces the benchmark energy at $\beta=1$, establishing a stable, physics-informed framework for finite-temperature quantum Monte Carlo where the nodal structure is learnt self-consistently.

\end{abstract}

\maketitle

\section{Introduction}\label{introduction}


The notorious Fermion sign problem appears in a number of guises, but everywhere we look it seems to enforce an \emph{exponential} asymptotic scaling of Fermionic quantum Monte Carlo.
Though there is a mathematical proof\cite{Troyer2005} that in general this problem is NP-hard, we may hope that physical Hamiltonians have more benign behaviour, and thus we can hope to win individual battles\cite{Zaanen2008}, if not the overall war.

The restricted path-integral Monte Carlo (RPIMC) method of Ceperley\cite{Ceperley1991,Ceperley1992} provides a \emph{polynomial} scaling finite-temperature quantum Monte Carlo framework, provided one knows the exact density matrix.
Using the nodal structure to restrict the paths to nodal regions, the fluctuating sign is sidestepped. 
As with other fixed-node quantum Monte Carlo methods, accuracy is ultimately limited by the quality of this nodal structure, which must be provided from elsewhere. 
This presents a paradox: acquiring an accurate many-body nodal surface requires having solved the quantum many-body problem, rendering subsequent Monte Carlo evaluation of observables redundant.
Even with the exact nodes, canonical path integral Monte Carlo can get stuck, and though still formally ergodic by the tiling theorem\cite{Ceperley1996Fermions}, practical convergence can be exponentially slowed.

The celebrated Worm algorithm of Boninsegni, Svistunov and Prokof'ev\cite{Boninsegni2006A,Boninsegni2006B} offers a general method to improve ergodicity of path integral methods by permitting disconnected path-integral world lines. 
The conintuous canonical path integral worldline separates into a `Worm' with head (Ira) and tail (Masha), named after two sisters, which are then freed to explore configuration space (formally we are sampling the Matsubara Green's functions, the G-sector) before reconnecting to the partition function (a canonical worldline path, the Z-sector).
Curiously, very little work has attempted to apply the Worm algorithm in a restricted path integral setting, except for the recent work of Fantoni\cite{Fantoni2021}.

Neural network wavefunctions\cite{Hermann2023} have enabled considerable recent progress in zero-temperature quantum Monte Carlo, but the method seems to struggle to learn the wavefunction as the number of Fermion particles increases.
Though practitioners of the method argue this is simply a matter of available computer time, perhaps this is just the exponential scaling of the sign problem appearing in a new guise, and that directly trying to learn the maximally entangled state is too hard.

Physical insight into quantum statistical mechanics motivated us to construct an alternative approach where the complexity of the quantum correlations can be learnt as they grow in thermodynamic time.  
At extremely high (thermonuclear) temperatures ($\beta = \frac{1}{k_B T} \to 0$), we reach a theoretical purity: kinetic energy dominates over Coulomb interactions and exchange, and the many-body density matrix converges to an exact, closed-form reference (the free-particle heat kernel). 
As the system cools towards zero, quantum statistical mechanics is a continuous projection in imaginary (thermodynamic) time ($e^{-\beta \hat H}$).  
By basing our machine-learning model directly in this projection, we we can model the quantum correlations as they grow with descending temperature, and then use our machine learnt model to accelerate the calculation at the next stage of the temperature curriculum. 
This should be easier than directly learning the full interacting ground state. 

At each step of the temperature curriculum, the learnt density matrix $\rho_\theta$ enhances the signal-to-noise of the restricted path-integral sampling (at this temperature), while the parameters $\theta$ provide a warm-start for the next, cooler temperature.
We evaluate the the error in the location of our Fermion nodes (i.e. our machine-learning loss) by the residual of the Bloch equation, the finite-temperature analogue of the local-energy in ground state variational Monte Carlo.

Our $\rho_\theta$ model for the density matrix is a Slater determinant, calculated from a set of backflow quasi-positions.
These quasi-positions are not from a closed-form equation, but from a continuous normalising flow (CNF), which learns the velocity field (driving an equation of motion) transforming  Fermion positions into quasi-positions.
The DeepSets approach ensures that the velocity model of the equation of motion is equivariant, which in turn enforces the correct Fermion quantum statistics in the determinant, at every point in training.

The motivation to define an equation of motion along imaginary time $t \in (0,\beta]$ for the backflow quasi-positions is that the ground state is highly structured and perhaps fractally self-similar\cite{Kruger2008}, but one might hope and expect that the driving force along continuous imaginary time generating this complexity is simple. 
Simple equations of motion are known to be capable of producing chaotic complexity and fractals.

The Slater determinant fully describes the nodal structure, but lacks the ability to express continuous variations, and led to ill-conditioned training. 
We therefore also learn a simple symmetric Jastrow factor (on top of a now \emph{invariant} DeepSets representation), which though not used in the restricted path integral Monte Carlo algorithm, stabilises training by regularising the loss landscape and enables us to more fully minimise the Bloch residual.

Lawrence and Yamauchi\cite{Lawrence2021} have shown that normalising flows can reduce the Fermionic sign problem in lattice models by deforming the path-integral contour. 
Luo and Clark\cite{Luo2019} developed neural-network backflow transformations to capture correlations in zero-temperature calculations, and Liu and Clark\cite{Liu2024Unifying} have illuminated the connections between general neural network wavefunction Ansatz and backflow. 

However the closest ancestor of our approach is the earlier Variational Density Matrix (VDM)
method by Militzer and Pollock\cite{MilitzerPollockPRE2000}, in which a Gaussian-orbital trial is
evolved from the free particle reference by minimising the integrated Bloch
residual, and the resulting nodes then used to restrict PIMC\cite{MilitzerCeperleyPRL2000,
MilitzerCeperleyPRE2001,militzer2000THESIS}. 
\textsc{Spindrift} is a neural VDM, with the fixed single-Gaussian replaced
with an equivariant flow for a backflow Slater determinant and an invariant
Jastrow network, trained on sampled configurations $(\R,\R',t)$. 
As well as being more expressive, our method improves on VDM (see
\cite[\S3.5]{militzer2000THESIS}) in two key ways:
1) our equivariant flow determinant remains exactly antisymmetric under all $N!$ exchanges during optimisation, 2) our Ansatz retains the exact density matrix property of being symmetric under $\R \leftrightarrow \R'$. 

Initial experiments on $N=3$ electrons in a $\beta=1$ 2D harmonic trap show that the method enables stable curriculum training along imaginary time.
The learnt velocity field significantly improves the nodal structure from the infinite-temperature (non-interacting free-particle) reference, reproducing the benchmark energy (i.e. leaving no fixed-node error).

\section{Method}

To calculate a Fermionic observable, in path integral Monte Carlo we typically run a Bosonic simulation\cite{Ceperley1995} (where probabilities are well defined and positive), and construct a reweighting Fermionic observable which picks up the alternating sign required by Fermionic antisymmetry.
For energy this looks like,

\begin{equation}
    \langle \hat{E} \rangle_F
    = \frac{\sum_{[p]} (-1)^{p_{\mathrm{ex}}} \, P_{[p]} \, E_{[p]}}
           {\sum_{[p]} (-1)^{p_{\mathrm{ex}}} \, P_{[p]}}
    \equiv \frac{\langle \sigma E \rangle_B}{\langle \sigma \rangle_B},
\end{equation}
where $P_{[p]}$ is the (positive, Bosonic) sampling weight of the
path-integral configuration in permutation family $[p]$, 
$E_{[p]}$ the energy estimator of this permutation family, 
and the alternating Fermionic reweighting factor $(-1)^{p_{\mathrm{ex}}}$
counts the $p_{\mathrm{ex}}$ pair exchanges of the permutation, of which there
are $N!$ for $N$ particles. 
Naive Monte Carlo sampling of this ratio estimator suffers from catastrophic sign error as $N$ increases.

\subsection{Restricted path integral Monte Carlo}

In 1991 Ceperley\cite{Ceperley1991} introduced the \emph{restricted} path integral Monte Carlo approach, where instead of resumming the Bosonic observable, the path itself is constrained to the regions of positive and negative density matrix.
Once thereby constrained with the \emph{a priori} information about the nodal structure, Fermionic Monte Carlo then converges in polynomial time.
The Fermionic nodal structure thereby defines the \emph{reach}\footnote{I do not know why it was called this, but I would like to think it due to Le Guin's use of the word both as the limits of geographic knowledge, but also of human agency, \emph{``But all this was mere tales of the Reach, which are always strange, and only my father gave it much thought.''} - Ursula K. Le Guin\cite{UrsulaLeGuin}.} of the restricted paths.

Here we implement restricted path-integrals on top of our \textsc{Halcyon.jl} Julia\cite{bezanson2017julia} path-integral Monte Carlo code, which in turn is essentially a faithful implementation of Spada et al.'s\cite{Spada2022} description of the Worm algorithm. 
Developing in Julia permits the heavily optimised, low-level, Worm Monte Carlo engine to sit in the same codebase and memory context as the high-level neural density matrix models, accelerating both execution and development and sidestepping the traditional two-language problem.

\subsection{Restricted Worm algorithm}

The celebrated Worm algorithm\cite{Boninsegni2006A,Boninsegni2006B} accelerates the convergence of path integral Monte Carlo by permitting disconnected worldlines.
These ``Worms'', with sister head \emph{Ira} and tail \emph{Masha}, are then freed to wander across the Matsubara Green's function G-sector, before reconnecting to the density matrix Z-sector.

Here we build on Spada et al.'s\cite{Spada2022} excellent description of the algorithm with improved periodic boundary conditions, and adopt their nomenclature.

Though the Worm algorithm is known to improve ergodic behaviour of path integral Monte Carlo, and getting stuck exploring the \emph{reach} was a known challenge of the restricted approach in canonical path integral Monte Carlo, only Fantoni\cite{Fantoni2021} appears to have attempted to combine the methods.

Now the definition of a path being restricted is that the segment remains in the same nodal region (density matrix sign) as the reference.
Defining $\R_{\mathrm{ref}}$ as the reference slice at imaginary time $0$, the restriction on path $\R_{\mathrm{ref}}$ to $\R_t$ at time $t$ is that
\begin{equation}
    \rho(\R_t,\R_{\mathrm{ref}};t) > 0
    \qquad \forall\, t\in(0,\beta].
    \label{eq:restriction}
\end{equation}
In the closed \textbf{Z-sector} (Monte Carlo moves \texttt{translate!}, \texttt{redraw!}, \texttt{close!}), we strictly enforce the nodal restriction.
When the Worm is open in the \textbf{G-sector} this requirement is relaxed, the open ends may therefore cross the trial nodal surface during \texttt{open!} and head/tail moves.
However, the closure slice $j=M$ dictates the overall permutation: $\R_M = \mathcal{P}\R_0$ for the permutation $\mathcal{P}$ encoded in the polymer topology. 
By the antisymmetry and positivity of the diagonal element $\rho_\theta(\R_0,\R_0;\beta)$, we then have
\begin{equation}
    \mathrm{sign}\,\rho_\theta(\mathcal{P}\R_0,\R_0;\beta)
    = (-1)^{\mathcal{P}}.
    \label{eq:closure-sign}
\end{equation}
Therefore, demanding positivity at $j=M$ upon closure acts precisely as a veto on odd permutations entering the Z-sector. 
Now, this means that open Worms will explore and repeatedly attempt to reconnect to the Z-sector, being rebuffed if this closure would cross a nodal boundary. 
The Worms feel the full interacting Hamiltonian, but if the density matrix is not exact then the nodal boundaries are in the wrong location, and the open-Worm (which is correctly sampling the thermal distribution of the interacting Hamiltonian) is continually rebuffed. 
For us, this serves as an additional diagnostic metric as we learn along the temperature curriculum: a large number of failure-to-close Worms indicates that we do not yet have a good model density matrix. 
(Fantoni\cite{Fantoni2021} noticed this trapping effect when using a free particle density matrix and therefore introduced nodal restrictions in their open Worm G-sector moves to ameliorate it.)

Overall, only minor, but subtle, additions are required to Spada et al.'s
Algorithm\cite{Spada2022} to adapt to a restricted Fermion formalism.

\subsection{Model density matrix}

In our algorithm, the density matrix $\rho$ is a learnt function with parameters $\theta$.

To enable temperature curriculum learning with a well-found starting point we initialise it with the correct high temperature reference (therefore it only needs to learn the quantum mechanical correlations as temperature drops).
In order to maximise the efficiency of learning we imbue the representation with the physical symmetries we know to be present.

In the `purity' of infinite temperature ($\beta=0$), quantum mechanics becomes trivial.
Kinetic-energy dominates ($\hat{T} \gg \hat{V}, \hat{U}$) and so the particles are independent, interactions are irrelevant, they simply diffuse.

The non-interacting single-particle imaginary-time propagator is the free-particle heat kernel,
\begin{equation}
    K_0(\rt,\rt';t)
    = \frac{1}{(4\pi\lambda t)^{D/2}}
      \exp\!\left(-\frac{|\rt-\rt'|^2}{4\lambda t}\right).
    \label{eq:free-kernel}
\end{equation}
Here and in the following, we have set $\hbar = 1$ and write the kinetic
coefficient as $\lambda \equiv \hbar^2/2m$; 
for the harmonic trap
$V(\rt)=\frac{1}{2}k|\rt|^2$ this gives $\omega = \sqrt{2\lambda k}$, and
our benchmark units $\lambda = \tfrac12$, $k = 1$ correspond to
$m = \omega = 1$ (oscillator units)

For $N$ spin-polarised Fermions, the corresponding many-body reference density matrix $\rho_0$ is simply the Slater determinant of single-particle kernels, 
\begin{equation}
    \rho_0(\R,\R';t) = \det\bigl[K_0(\rt_i,\rt'_j;t)\bigr]_{i,j=1}^{N}.
\end{equation}

This is our free-particle reference, exact at $t=0$, from which we then learn the appearance of quantum correlations in the nodal surface as temperature drops $t \to \beta$.

\subsubsection{Continuous normalising flow Slater backflow}

Interactions and exchange deform the nodal surface away from the free-particle reference.
Inspired by previous work\cite{Kruger2008,Zaanen2008} showing that surprisingly complex wavefunctions can be calculated from determinants of simple analytic backflows, we learn a mapping to quasi-position by integrating a learnt velocity field, a continuous normalising flow.

This maps from electron positions $\R$ to backflow quasi-positions $\Rt$ by integration of an ordinary differential equation along a fictitious time-like variable, $s\in[0,1]$,

\begin{equation}
    \frac{d\R_s}{ds} = t\,\bm{v}_\theta(\R_s,s,t),
    \qquad
    \Rt \equiv \R_{s=1}
    = \R + t\int_0^1 \bm{v}_\theta(\R_s,s,t)\,ds.
    \label{eq:cnf-ode}
\end{equation}
The velocity field is permutation equivariant,
$\bm{v}_\theta(\mathcal{P}\R,s,t)=\mathcal{P}\,\bm{v}_\theta(\R,s,t)$ (a DeepSets\cite{zaheer2017deep} construction), so relabelling particles relabels their velocities and nothing more.

The factor of $t$ (imaginary time) is key.
Therefore $\lim_{t\to0}\Rt = \R$ and the exact high-temperature density matrix limit is recovered smoothly and automatically.
This additionally imposes an (approximate) structure on the curriculum learning that the degree of backflow is proportional to imaginary time.

These quasi-particle locations are then directly used in a Slater determinant,
\begin{equation}
    S_\theta(\R,\R';t)
    = \det\bigl[K_0(\rtt_i,\rtt'_j;t)\bigr]_{i,j=1}^{N}.
    \label{eq:slater}
\end{equation}
Fermionic antisymmetry is strictly maintained. 
Because the flow is equivariant, exchanging two particles in $\R$ exchanges the corresponding blocks of $\Rt$. 
This swaps two rows of the kernel matrix $K(\rtt_i,\rtt'_j;t)$ and multiplies the determinant by $-1$.
This holds for any learnt velocity field $\bm{v}_\theta$.

Now, a perceived limitation might be that integrating over this velocity field smoothly deforms the configuration space, which is then preserved through the Slater determinant, with no ability to change topology (this is the defining diffeomorphism property of CNFs). 
Yet we know that abrupt phase transitions (condensation of a plasma, quantum degeneracy, macroscopic quantum phases) occur along imaginary time. 
Ceperley's analysis of Fermion nodal properties in continuous spatial dimensions $D \ge 2$ and the associated tiling theorem\cite{Ceperley1991} demonstrate that the spin-polarised Fermion ground state partitions configuration space into exactly two maximally connected pockets. 
Because our high-temperature, non-interacting reference possesses this identical two-pocket topology, a smooth topological deformation exists connecting the free-particle reference all the way to the exact ground state. 
It is precisely this continuous deformation that \textsc{Spindrift} seeks to learn along the temperature curriculum. 

While the \emph{topology} of the two nodal regions thus remains invariant, the \emph{geometry} can become extraordinarily convoluted as correlation condenses. 
Kruger and Zaanen\cite{Kruger2008,Zaanen2008} hypothesise that the nodal surfaces of strongly correlated Fermions can exhibit fractal character at quantum critical points, developing intricate, self-similar folding across all length scales, which they model with a Feynman-Cohen analytic backflow. 
As our continuous normalising flow has a finite maximal structure, it cannot support a true fractal, but instead we are effectively limited by this cut-off to generate a pseudo-fractal (i.e. multiscale, folded, complexity at intermediate scales, but then being strictly smooth in the limit of infinitesimal scale). 
One extension to our method would be to train our model by flow-matching to analytic Feynman-Cohen backflows, and then develop and benchmark dynamic fractal analysis (e.g. finite-time Lyapunov exponents of the continuous normalising flow) to the static Hausdorff dimensions reported by Kruger and Zaanen. 
We would then be in a position to go hunting for these fractal nodal structures arising directly from a \text{Spindrift} simulation of real materials.

We note that extending our method to unpolarised Fermions will require additional topological considerations. 
Mitas\cite{Mitas2006PRL,Mitas2006ArXiv} rigorously proved that while the spin-polarised nodal topology is invariant, arbitrarily weak interactions in unpolarised systems collapse the four nodal regions of the non-interacting system down to just two, a topological change.


\subsubsection{Jastrow-net}

While the Slater determinant carries the nodal structure, which is strictly all that is needed to \emph{evaluate} restricted path integrals, we found that learning just this was ill-conditioned. 
This determinant cannot express the smooth modulation of the density matrix between nodes. 
To regularise the loss landscape, we adopt a standard technique from the quantum Monte Carlo community, augmenting our backflow transformation with a symmetric Jastrow factor. 

The full composite model density matrix is position-symmetric and permutation-invariant,
\begin{equation}
    \rho_\theta(\R,\R';t)
    = S_\theta(\R,\R';t)\,
      \exp\!\bigl[-J_\theta(\R,\R';t)\bigr],
    \label{eq:full-ansatz}
\end{equation}
with
\begin{equation}
    J_\theta(\R,\R';t)
    = t\,\bigl[u_\theta(\R,t) + u_\theta(\R',t)\bigr],
    \label{eq:jastrow}
\end{equation}
where $u_\theta$ is a DeepSets\cite{zaheer2017deep} \emph{invariant} scalar network.

This is \emph{invariant} (in contrast to the \emph{equivariant} continuous normalising flows), because an amplitude must not care about particle labels at all.
By construction this Jastrow factor is strictly positive, so it cannot move the nodes, and encodes the spatial symmetry $\R\leftrightarrow\R'$ required of a density matrix.

As with the backflow, the explicit factor of $t$ (imaginary time) guarantees that the (exact) free-particle density matrix is smoothly recovered.
Practically, the output layer of $u_\theta$ is zero-initialised, so $J_\theta\equiv0$. 
This makes a stable and unbiased starting point (free particle) for curriculum training.

Our chosen form of Jastrow was deliberately simple, and \emph{separable} (containing no terms that couple $\R$ and $\R'$). 
Therefore it can only modulate one-body amplitudes, and cannot express cross-correlations (e.g. the Coulomb cusp). 
This does not affect the \emph{restricted} PIMC sampling (which depends only on the node locations), but this likely explains why the Bloch residual cannot currently be optimised to zero. 
We return to this point when considering our interacting experiments below.

In future work we could incorporate a pairwise Jastrow, though care is necessary to respect the required symmetry.
Our reference could also be improved with Pollock's exact finite-temperature Coulomb cusp form of the Jastrow factor\cite[Appendix~B]{militzer2000THESIS}. 

\subsection{Local energy and the Bloch residual}

How do we know when the learnt density matrix is \emph{wrong}, without access to the exact answer?
The exact density matrix satisfies the (partial differential) Bloch equation,

\begin{equation}
    -\frac{\partial\rho}{\partial t} = \mathcal{H}\rho
    = -\lambda\,\nabla_{\R}^2\rho + V(\R)\,\rho.
    \label{eq:bloch}
\end{equation}

This can be considered the finite-temperature (imaginary-time) equivalent to the Schr\"odinger equation, and is of the form of a heat equation with a potential-dependent sink.
Physically, as we integrate along imaginary time, the amplitude diffuses (kinetic term) and decays where the potential is high.

A somewhat surprising fact is that any candidate density matrix can be tested against this equation pointwise.
Following Ceperley\cite{Ceperley1995}, we define the local energy as:

\begin{equation}
    E_L(\R,\R';t)
    \equiv \frac{1}{\rho} \mathcal{H}\rho
    = -\frac{\lambda}{\rho}\,\nabla_{\R}^2\rho + V(\R),
    \label{eq:local-energy}
\end{equation}
and the corresponding pointwise Bloch residual as:
\begin{equation}
    E_{\mathrm{res}}(\R,\R';t)
    \equiv \frac{1}{\rho}\left[\frac{\partial}{\partial t} + \mathcal{H}\right]\rho
    = \frac{\partial}{\partial t} \ln \rho + E_L(\R,\R';t),
    \label{eq:bloch-residual}
\end{equation}
so $E_{\mathrm{res}} \equiv 0$ for the exact $\rho$.

This is the finite-temperature analogue of the `local energy' from ground-state variational Monte Carlo.
We assume the same zero-variance principle: for the exact $\rho$ the Bloch residual should vanish identically, at \emph{every} sampled point, not just the integral.
If we find our learned parameters $\theta$ such that $\rho_\theta$ is exact, this residual drops to zero.
We can therefore hope to use the residual directly as the loss in our machine learning.

The thing we care most about is the location of the nodes.
On the trial nodal surface $\rho_\theta(\R_0,\R';t)=0$, the potential term is multiplied by zero and drops out from the unnormalised residual operator $\left[\frac{\partial}{\partial t} + \mathcal{H}\right]\rho_\theta$:
\begin{equation}
    \left[\frac{\partial}{\partial t} + \mathcal{H}\right]\rho_\theta\bigg|_{\R=\R_0}
    = \left.\frac{\partial\rho_\theta}{\partial t}\right|_{\R_0}
      - \left.\lambda\,\nabla_{\R}^2\rho_\theta\right|_{\R_0}.
    \label{eq:node-operator}
\end{equation}

A misplaced trial node breaks this balance, and the resulting non-zero value is the gradient signal that pushes $\theta$ to move the zero locus: the Bloch equation itself tells us where the nodes should flow.

\subsubsection{Logarithmic variance near nodes}

A naive application of a local energy loss was unstable during training; simple systems would seem to improve and then suddenly blow up.
Again, much brute-force effort was spent attempting to regularise this.
Near a node at perpendicular distance $x$, $\rho_\theta\sim x\,g$ with $g=|\nabla_{\R}\rho_\theta|$.
If $\left[\frac{\partial}{\partial t} + \mathcal{H}\right]\rho_\theta \sim \mathcal{E}_0\,g$ at the node, then $E_{\mathrm{res}} \sim \mathcal{E}_0/x$: the Bloch residual blows up as $1/x$ on approach.

Initially we could not understand why this was not a general problem: the same singularity is present in ground-state variational Monte Carlo (VMC).
The answer appears to be that in VMC, the the sampling weight $|\Psi_T|^2\propto x^2$ vanishes \emph{quadratically}
at the nodes, so the integral $\int_0^\delta x^2(1/x)^2\,\mathrm{d}x$ remains finite. 
(This quadratic suppression separately leads to the apparent paradox that the
physically decisive nodal regions are massively undersampled in
VMC\footnote{Ali Alavi, private communication over Pizza, April 2026.}.)
We are not so lucky!
Under the restricted pair measure (Equation \eqref{eq:pair-measure}) only the factor $\rho(\R,\R';t)$ vanishes linearly in $x$; the complementary factor $\rho(\R',\R;\beta-t)$ is evaluated at a different imaginary time where we can expect it to be non-zero.

Hence $P\propto x$ only, and
\begin{equation}
    \mathrm{Var}_P\bigl[E_L\bigr]
    \propto \int_0^\delta x\left(\frac{1}{x}\right)^2\mathrm{d}x
    = \infty.
    \label{eq:log-divergence}
\end{equation}
This divergence is `only' logarithmic, but it means occasional near-node samples
dominate the loss with essentially unbounded values, producing the observed instability.

Having now understood the origin of this divergence which entered into our loss, we tamed it by empirically defining a boundary length $\tau_{\mathrm{r}}$ near a node.
A natural way to define distance to a node is $|\nabla_{\R}\rho_\theta|/|\rho_\theta|\sim 1/x$.

This we can then regularise with
\begin{align}
E_{\mathrm{res},\mathrm{reg}}(\R,\R';t)
    &= \frac{E_{\mathrm{res}}}
           {1 + \tau_{\mathrm{r}}^{2}\,|\nabla_{\R}\rho_\theta|^2/\rho_\theta^2} \nonumber \\
    &= \frac{\bigl( \left[\frac{\partial}{\partial t} + \mathcal{H}\right]\rho_\theta \bigr)\,\rho_\theta}
           {\rho_\theta^2 + \tau_{\mathrm{r}}^{2}\,|\nabla_{\R}\rho_\theta|^2}.
    \label{eq:r-reg}
\end{align}
Away from a node the denominator is essentially unity.
Near a node, substituting $\rho_\theta\approx xg$ and $|\nabla_{\R}\rho_\theta|\approx g$ cancels the amplitude $g$ entirely,
\begin{equation}
    E_{\mathrm{res},\mathrm{reg}}(x)
    \approx \mathcal{E}_0\,\frac{x}{x^2+\tau_{\mathrm{r}}^2}.
    \label{eq:node-reg}
\end{equation}
Rather than diverging, the integral has a maximum at the boundary
$x=\tau_{\mathrm{r}}$, and is smoothly pushed to zero while retaining the
correct direction of diffusion, and preserving the solution---the global
minimum of the loss remains guiding the solution towards the exact density
matrix. 

Though this was developed in a rather ad-hoc manner, it appears that the
quantity $\rho_\theta/|\nabla_{\R}\rho_\theta|$ is the same Newton-Raphson
estimate of the distance to the nodal surface used by
Militzer to construct the nodal action\cite[Eq.~2.94]{militzer2000THESIS}. 
From a machine learning point of view, this physically motivated boundary layer could be understand as the Huber loss used in robust regression, to avoid domination by outliers. 
There are almost certainly better and more well-motivated approaches to regularise this divergence. 

Currently $\tau_{\mathrm{r}}$ is treated as a constant hyperparameter, but clearly this could be scaled or adapted as the simulation progresses. 

This regularised Bloch residual is then used with standard linear gradient-based optimisers and back propagation to train the machine learning models.

\subsubsection{Stochastic estimators}

Evaluating the residual requires $\partial_t\rho_\theta$ and the many-body Laplacian $\nabla_{\R}^2\rho_\theta$.
Initially we attempted to automatically differentiate these objects as part of the training loop, but found this both computationally expensive and challenging to work reliably.
Therefore we fall back to finite differences.

With finite-difference step $h$, we have

\begin{align}
    \frac{1}{\rho_\theta}\frac{\partial\rho_\theta}{\partial t}
    &\approx \frac{r_{t+h}-r_{t-h}}{2h},
    \label{eq:dt-fd}
    \\
    \frac{1}{\rho_\theta}\,\nabla_{\R}^2\rho_\theta
    &\approx \left\langle\frac{r_{\R+hu}-2+r_{\R-hu}}{h^2}\right\rangle_{u},
    \label{eq:lap-fd}
\end{align}
where $r_x\equiv\rho_\theta(x)/\rho_\theta(0)$ is the ratio of the shifted to the centred evaluation.
As an aside, this is all done in log-differences to avoid numeric underflow.
The full Laplacian in $ND$ dimensions would cost $2ND$ evaluations per sample;
alternatively the bracket $\langle\cdot\rangle_u$ is the \emph{mean} over a set of random Rademacher probe directions $u$.
Care is then needed: because the loss is the square of the residual, a noisy estimate can lead to bias. 

In the regularised local energy\eqref{eq:r-reg} we also need the gradient \emph{norm} $|\nabla_{\R}\rho_\theta|^2$.
\begin{equation}
    \mathbb{E}_u\bigl[(u\cdot\nabla_{\R}\rho_\theta)^2\bigr]
    = \sum_{i,j}\partial_i\rho_\theta\,\partial_j\rho_\theta\,\mathbb{E}[u_iu_j]
    = |\nabla_{\R}\rho_\theta|^2.
    \label{eq:grad-identity}
\end{equation}
The central difference approximates the directional derivative to $O(h^2)$,
\begin{align}
    \frac{r_{\R+hu}-r_{\R-hu}}{2h}
    &= \frac{u\cdot\nabla_{\R}\rho_\theta}{\rho_\theta} + O(h^2), \\
    \mathbb{E}_u\!\left[
      \left(\frac{r_{\R+hu}-r_{\R-hu}}{2h}\right)^{\!2}
    \right]
    &\approx \frac{|\nabla_{\R}\rho_\theta|^2}{\rho_\theta^2}.
    \label{eq:grad-estimator}
\end{align}
This requires no additional density matrix evaluations.

Combining \eqref{eq:dt-fd}--\eqref{eq:grad-estimator} with the potential $V(\textbf{R})$ and background offset $E_{\mathrm{ref}}$, evaluated stochastically (over $u$ Rademacher probes) is
\begin{widetext}
\begin{equation}
    E_{\mathrm{res},\mathrm{reg}}
    = \frac{
        \dfrac{r_{t+h}-r_{t-h}}{2h}
        - \lambda\left\langle\dfrac{r_{\R+hu}-2+r_{\R-hu}}{h^2}\right\rangle_u
        + V(\R)
      }{1 + \tau_{\mathrm{r}}^2\left\langle\left(\dfrac{r_{\R+hu}-r_{\R-hu}}{2h}\right)^{\!2}\right\rangle_u} .
    \label{eq:r-reg-assembled}
\end{equation}

\end{widetext}

Our current learning loss is simply the square of the above $L=E_{\mathrm{res},\mathrm{reg}}^2$.

\subsubsection{Training data}

To evaluate the above Bloch equation residual energy, we need pairs of configurations with known imaginary-time separation drawn from the physical distribution.
This is exactly what we have in the (restricted) path integral Monte Carlo state vectors.
In contrast to variational Monte Carlo, the diffuse paths sample up against the nodal regions.

Any two slices of an equilibrated thermal loop, separated by segment length $t$, should be distributed according to the pair density
\begin{equation}
    P(\R,\R';t)
    = \frac{1}{Z}\,
      \rho_F(\R,\R';t)\,\rho_F(\R',\R;\beta-t),
    \label{eq:pair-measure}
\end{equation}
this is just the loop probability marginalised onto the two chosen slices: one factor for the direct route, one for the return.

We draw these $(\R,\R',t)$ from Z-sector Worm configurations: a random origin slice $j_0$, segment $\Delta j\in\{1,\ldots,\lfloor T_{\max}/\tau\rfloor\}$ (where $\tau = \beta/M$ is the discrete imaginary time step), and
\begin{equation}
    t = \Delta j\,\tau,
    \qquad
    \R = \R_{j_0},
    \qquad
    \R' = \R_{j_0+\Delta j}
    \label{eq:segment-sample}
\end{equation}
(with periodic wrapping along the polymer worldline). 
Two caveats---these samples are by \emph{restricted} PIMC under the current
trial nodes, so $P$ is a fixed-node trial measure and only exact when
$\rho_\theta$ is exact (which it is not); and each curriculum stage lags the
previous one, so one hopes the self-consistency bias vanishes in much the same
way as an iterative mean-field method. 

\subsubsection{Curriculum learning along imaginary time}

A key motivation of this work was to find a mechanism by which we could learn the appearance of quantum correlations continuously, as the path-integral Monte Carlo algorithm projects down from the independent (infinite temperature; imaginary time $t=0$) starting configuration.

Therefore, we train the model on a curriculum, with increasing $\beta$.
Initially the $t\to0$ free-particle reference (no backflow, $\lim_{t\to0}\Rt = \R$) provides the exact free-particle reference, and then the first stage only has to learn the relatively simple nodal structure which develops at high temperature.

Each temperature stage then inherits the previous stage's model $\theta$ as a warm-start to the learning, and also benefits from the learnt $\rho_\theta$ in improving the sampling efficiency of the restricted path integral Monte Carlo. 

Practically speaking, we have observed that the Worm algorithm actually provides a direct metric of whether the $\rho_\theta$ is being accurately modelled. 
When learning is incomplete or diverging, we find that the algorithm gets increasingly trapped in the $G$-sector, as attempts to reconnect to $Z$ are rejected because the locations explored by the Worm (following the true Hamiltonian) no longer align with a poor model for the nodal structure.

\subsubsection{On-the-fly mixed thermodynamic estimator}

By rearranging the Bloch equation (Eq.~\ref{eq:bloch}), the spatial local energy must equal the logarithmic imaginary-time derivative. 
Evaluating this derivative for our trial ansatz $\rho_\theta$ over the exact thermal pair-measure $P$ samples (Eq.~\ref{eq:pair-measure}) provides the on-the-fly mixed thermodynamic estimator, 
\begin{equation}
    E_{\mathrm{mixed}} \equiv \left\langle -\frac{\partial}{\partial t} \ln \rho_\theta(\R,\R';t) \right\rangle_{P}.
\end{equation}
We call this ``mixed'' because we combine exact path-integral samples with the model density matrix. 
Practically, we found this a very stable energy estimator, and monitor the
median (which resists spikes from the positive kinetic-energy divergence in the near-node samples) on-the-fly as the model trains. 
As the Bloch residual is minimised by the Zero-Variance principle, the variance of this estimator collapses.
Using the \emph{median} systematically discards the high-energy tail, and thus appears to underestimate the true mean thermodynamic energy. 

\section{Results}

\subsection{A finite temperature Fermion in a harmonic trap}

As happens with disturbing regularity in modern physics, we once again find ourselves studying the harmonic oscillator.

For a single particle of mass $m$ in a 1D harmonic oscillator potential ($V(x)=\frac{1}{2} m \omega^2 x^2$), the exact single-particle imaginary-time propagator is known in closed form (the Mehler formula),
\begin{widetext}
\begin{equation}
    K_{\mathrm{HO}}(r,r';t) = \sqrt{\frac{m\omega}{2\pi \sinh(\omega t)}} \exp\left[ -\frac{m\omega}{2\sinh(\omega t)} \left( (r^2+r'^2)\cosh(\omega t) - 2rr' \right) \right] .
    \label{eq:mehler-kernel}
\end{equation}
\end{widetext}
This is an exact solution to the Bloch equation.

At extremely high-temperature, $t \to 0$, $\sinh \omega t \approx \omega t$ and $\cosh \omega t \approx 1$, and so we can simplify to,

\begin{equation}
    K_{\mathrm{HO}}(r,r';t \to 0) \approx \sqrt{\frac{m}{2\pi t}} \exp\left[ -\frac{m}{2t} (r-r')^2 \right] \equiv K_0(r,r';t).
\end{equation}

This is the free-particle limit. 
In higher dimensions, the dimensions separate so the overall density matrix is the product. 

These exact solutions were key to debugging the code around evaluating the Bloch residual.

\subsection{Non-interacting Fermions in a 2D harmonic trap}

For $N$ non-interacting Fermions, we can directly construct the $N$-particle density matrix $\rho_F$ by taking a Slater determinant of the single-particle Mehler kernels,
\begin{equation}
    \rho_F(\R,\R';t) = \det \bigl[ K_{\mathrm{HO}}(\rt_i,\rt'_j; t) \bigr]_{i,j=1}^N.
    \label{eq:mehler-slater}
\end{equation}
This realisation was essential for development of this method.
Combined with the $t \to {0, \infty}$ limits above, it developed understanding of the physics.
Inserting this density matrix into the Bloch residual evaluates to zero everywhere as it exactly solves the Bloch equation (this fact was used to fix several bugs in our codes!).
This exact reference was also useful to debug the equivariance and invariance of the CNF and Jastrow models, and correctly form the reference density matrix.

The Mehler kernel has a simple physical interpretation - it flows the Fermions
towards the centre of the trap, feeling the effect of the external potential.
The nodal structure of the Mehler kernel comes purely from the Pauli exclusion principle, mathematically expressed as the Slater determinant, which generates a pure exchange-hole around each particle with a size proportional to the thermal de Broglie wavelength in the harmonic trap (the Mehler kernel above, which becomes bounded by the harmonic oscillator length scale $l = \sqrt{\hbar / m\omega}$). 

Running \textsc{Spindrift} on this non-interacting system should pick up the same physics as it learns $K_0 \to K_{\mathrm{HO}}$: the
Jastrow factor should pick up the effect of the external potential, and the
velocity field in the continuous normalising flow should enforce the Pauli
exclusion.

\subsection{Interacting Fermions in a 2D harmonic trap}

We benchmark our method on a system of interacting Fermions in a two-dimensional harmonic trap.
Specifically, we consider $N=3$ spin-polarised Fermions confined by a harmonic potential $V(\R) = \frac{1}{2} k \sum_{i=1}^N |\rt_i|^2$ with $k=1.0$.
The inter-particle interactions are modelled by a regularised Kelbg-Coulomb potential, preventing singularities at the origin.
We descend to a thermodynamic temperature of $\beta=1.0$.

In the canonical PIMC, the sign problem presents as a vanishing average sign $\bar\sigma \to 0$, whereas in our restricted Worm algorithm it presents as trapping in the $G$-sector, as invalid nodal topologies prevent Worm closure to the $Z$-sector. 
For the trained model, the $G$-sector trapping provides a direct metric of the nodal quality: at early intermediate curriculum stages the Worms struggled to reconnect ($T_{\mathrm{max}}=0.4$, diagnostic tunnelling count $\sim 521$).
As the Bloch residual spatial variance continues to decrease, the nodal surface became closer to the true interacting Hamiltonian, and this topological trapping continued to decrease (diagnostic tunnelling count $\sim 7$ at $\beta=1.0$).

\subsubsection{Training and hyper parameters}

The path integral is discretised with $M=200$ beads (time slices).

We train the continuous normalising flow model using $10,000$ gradient descent steps per curriculum stage, with a batch size of $128$ path segments drawn from the equilibrium distribution, at that temperature stage using the restricted Worm algorithm with the density matrix of the previous stage (and free particle density matrix to begin).
We use the standard Adam optimiser with a learning rate of $10^{-4}$ and weight decay of $10^{-5}$.
This is extremely slow (and rather arduous!) learning. 
Adam is a first-order optimiser, and our loss is a stiff PDE, and our small networks might be suffering in an unhealthy loss landscape. 
In Ferminet\cite{Pfau2020}, they found that they needed a carefully tuned KFAC approximate natural gradient optimiser. 
Future work will look at optimising this training. 

The boundary-layer regularisation parameter is set to $\tau_{\mathrm{r}} = 0.1$, and a small flow penalty of $10^{-3}$ is applied to keep the backflow bounded. 
To estimate the Laplacian in the Bloch equation, we use an exact finite difference estimator with a step size of $h=10^{-3}$. 
(We have also implemented a stochastic Hutchinson trace estimator for scaling to large systems, but do not use it here to avoid noise-induced bias.)
The continuous normalising flow is integrated using a Runge-Kutta 4 scheme with step size $\mathrm{d}s=0.2$.

Because we learn a \emph{correction} from a physical prior, our neural networks are extremely small; we learn the vector field perturbation from the free particle heat kernel (the CNF backflow, $\sim 3,300$ parameters), and a modulating amplitude (the Jastrow factor, $\sim 2,300$ parameters).

\subsubsection{Curriculum Learning and Nodal Evolution}
The core of the Spindrift algorithm is the continuous normalising flow that learns the nodal structure along the projection in imaginary time.

\begin{figure*}[htbp]
    \centering
    \includegraphics[width=1.0\textwidth]{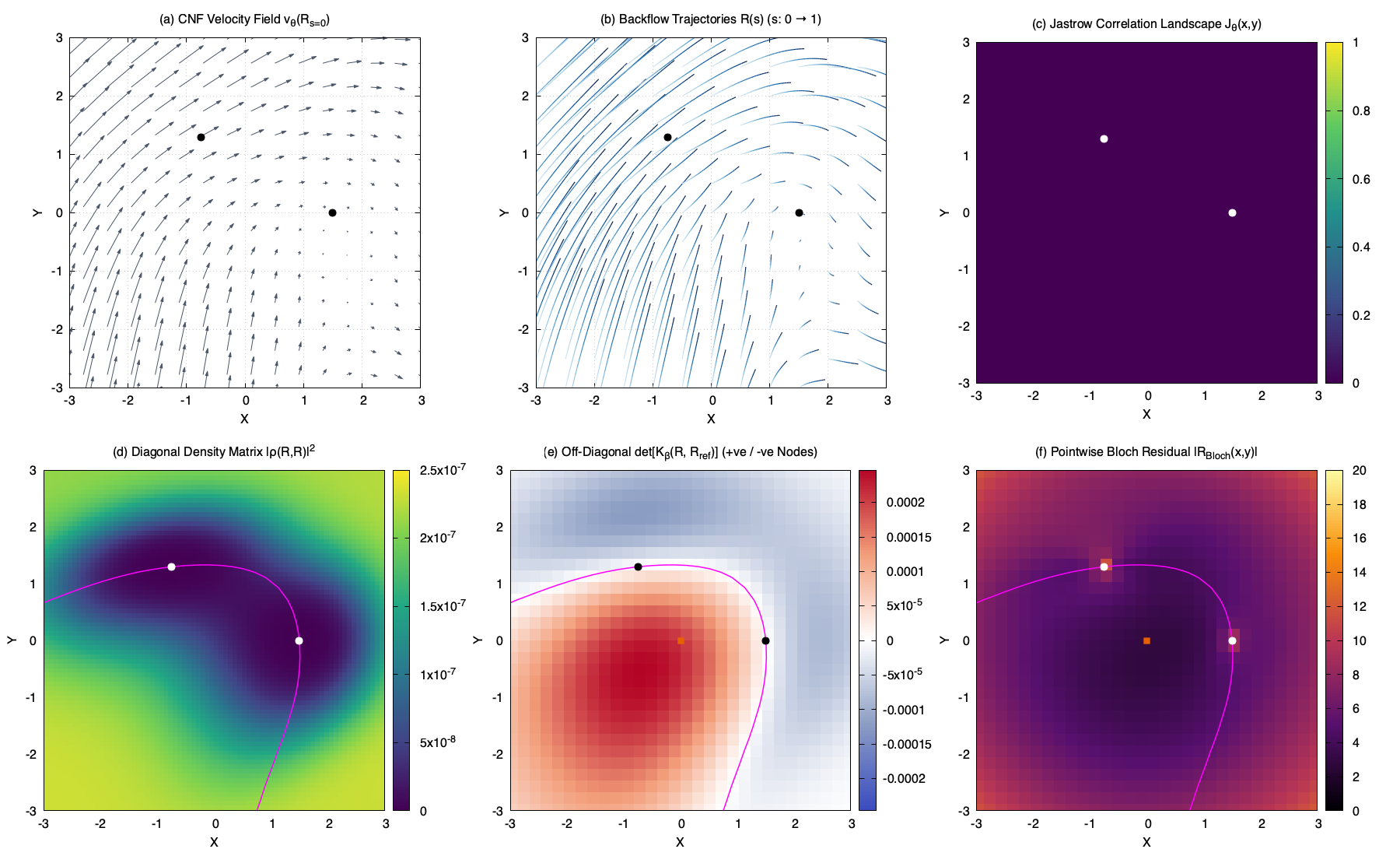}
    \caption{
    Initial state of the neural density matrix model prior to training, internal structure (top) and physical observables (bottom), for $N=3$ spin-polarised Fermions in a 2D harmonic trap. 
    Most figures are generated by constraining two electrons to sit on the Wigner radius ($r=1.5$, where the Coulomb and Harmonic force would balance), and then using the third electron as the test particle. 
    (a) The randomly initialised continuous normalising flow (CNF) velocity field $\mathbf{v}_\theta(R, s=0)$ shows vortices. 
    This is due to the permutation-equivariance enforced on the network; the random vector fields are thereby symmetric around the particle positions.
    (b) The random field generates rather large backflow trajectories.
    (c) The Jastrow is initialised as exactly zero across all space. 
    (d) The diagonal probability density reflects only Pauli exclusion, showing no preference for the harmonic trap centre.
    (e) Yet we still have nodal structure (magenta) in the off-diagonal Slater determinant.
    (f) The Bloch residual is a perfect image of the underlying harmonic trap potential and the Coulomb repulsion around the other electrons - the initial density matrix is the exact solution to the free-particle Bloch equation, and so the missing contribution in the Hamiltonian is simply the potential.
    }
    \label{fig:6panel_initial}
\end{figure*}

\begin{figure*}[htbp]
    \centering
    \includegraphics[width=1.0\textwidth]{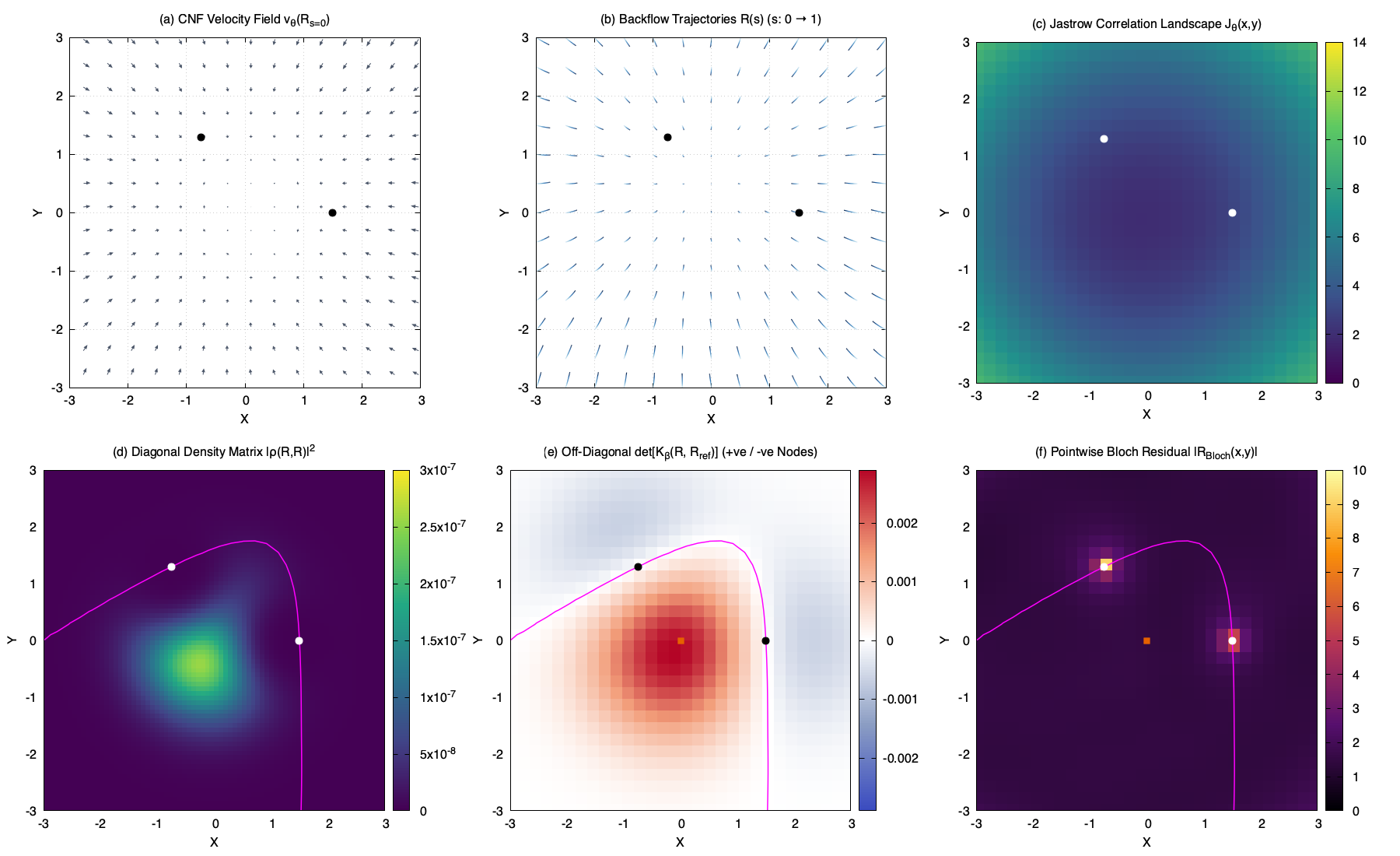}
    \caption{
    Neural density matrix model internal structure (top) and physical observables (bottom), for $N=3$ spin-polarised Fermions in a 2D harmonic trap at $\beta=1.0$. 
    (a) Continuous normalising flow (CNF) velocity field $\mathbf{v}_\theta(R, s=0)$ driving the quasi-particle transform. 
    (b) Integrated backflow trajectories $R(s)$ over fictitious time $s \in [0,1]$. 
    (c) Symmetric Jastrow correlation landscape $J_\theta(x,y)$. Clearly this has learnt the harmonic confinement of the trap, compared to the free-particle starting point. 
    (d) Diagonal probability density $|\rho(R,R)|^2$, showing spatial localisation of the 3rd electron. Pink contour is the nodal structure from the following panel. Clearly the model has learned to fill the third location in the Wigner polygon. 
    (e) Off-diagonal nodal domains via the two-point Slater determinant $\det[K_\beta(R, R_{\text{ref}})]$. The zero-contour line defines the nodal structure (magenta).
    (f) Pointwise local Bloch residual $|R_{\text{Bloch}}(x,y)|$, which defines the local error in the model density matrix. 
    Note that training reduces the pointwise Bloch residual roughly fourfold
    relative to the initial model displayed in 
Fig.~\ref{fig:6panel_initial}. 
The main remaining residual are the Coulomb cusps centered on the other electrons. 
    }
    \label{fig:6panel_trained}
\end{figure*}

In Figure \ref{fig:6panel_trained}, we present a sequence of two-dimensional heat-maps of internal parameters of the model, and physical observables, evaluated at $\beta = 1.0$. 
These are perhaps best understood in comparison to the initial (infinite
temperature free-particle reference, but evaluated at $\beta = 1.0$) model shown
in Figure \ref{fig:6panel_initial}. 
These figures could be compared to the zero-temperature wavefunction analogue inferred from Ferminet style architecture and VMC, as done by Freitas et al.\cite{Freitas2025} for the 20-electron configuration. 

The curriculum progresses from a high-temperature non-interacting reference
($T=0.2$) down to the target thermodynamic temperature ($T=\beta=1.0$).
As the curriculum drops in temperature, the continuous normalising flow
progressively pushes the non-interacting nodes, visibly deforming the nodal
pockets to produce a recognisable exchange-correlation hole for particles in
a harmonic potential.

\subsubsection{Learning curves}
\begin{figure}[htbp]
    \centering
    \includegraphics[width=\columnwidth]{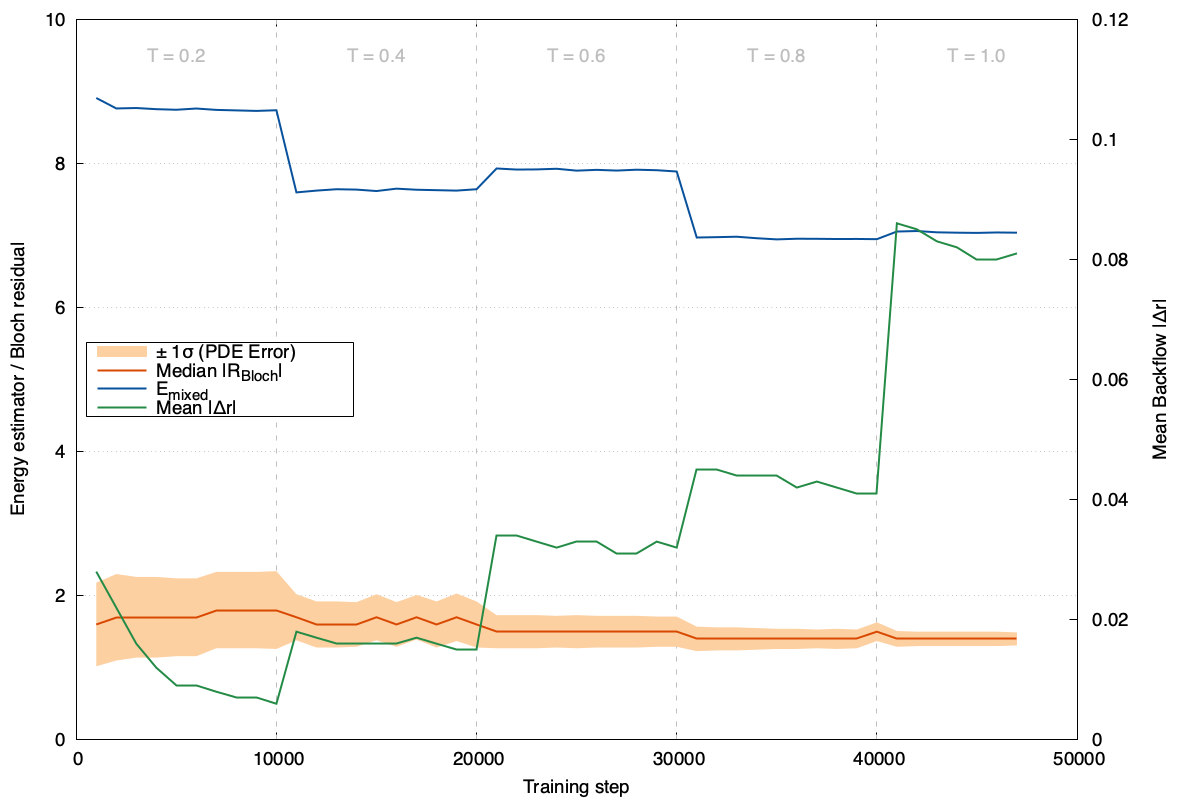}
    \caption{
        Training along the temperature curriculum ($\beta_{\mathrm{stage}}
        = 0.2, 0.4, \ldots, 1.0$, dashed boundaries). 
        Shown are the mixed energy estimator (blue, in units of $\hbar\omega$),
        the median Bloch residual energy with its spatial standard-deviation
        (orange), and the mean backflow displacement magnitude (green, axis on
        right). 
    As we approach $\beta=1$, we are entering quantum degeneracy, leading to the large up-tick in backflow strength to enforce the Pauli exclusion. 
    We can see that \textsc{Spindrift} successfully reduces the spatial
    standard deviation of the Bloch residual, while the median Bloch residual
    plateaus near $1\, \hbar \omega$. 
    This may indicate that while the algorithm and the learning are functioning, our current combination of Slater and Jastrow factors are not sufficiently expressive. 
    The final stage is shorter as the run was wall-clock limited: Vitka Korneev needed the machine back. 
    }
    \label{fig:learning_curvse}
\end{figure}

One significant challenge we faced in learning the density matrix from the Bloch equation residual is the logarithmic variance divergence near the nodal boundaries, $\mathcal{O}(1/x)$.
Figure \ref{fig:learning_curvse} demonstrates that our boundary-layer regularisation scheme successfully tames this divergence; the median residual exhibits a stable, monotonic descent over the training epochs for each stage of the curriculum, with a spatial standard deviation that continues to contract, confirming the robustness of the regularised loss function. 
However, the fact that it stabilises at around $1 \hbar\omega$ may indicate that our density matrix model is not expressive enough to fully learn the physics of the system. 

Similarly, it is at the final $\beta=1.0$ temperature curriculum that the physics starts to get interesting as the system becomes quantum degenerate, with a large up-tick in the required backflow strength. 

\subsubsection{Final energies}

Because we do not yet include the nodal-action correction (an effective
repulsion when the path approaches a node), we expected our \emph{restricted} estimators to systematically underestimate the kinetic energy. 
Dornheim provides a benchmark reference energy of $E \approx 8.719(3)$ for the same $N=3$ $\beta=1$ system. 

We first ran our restricted Worm algorithm using the free-particle nodes ($1\times10^5$ Monte Carlo steps). 
The thermodynamic $E_{\mathrm{thermo}} \approx 7.242$ and virial $E_{\mathrm{virial}} \approx 8.316$ estimators strongly disagree and both severely underestimate the reference. 

Monitoring our biased ``mixed'' estimator, the energy of the learnt nodal surface lies below the free-particle reference at
\emph{every} curriculum temperature (Fig.~\ref{fig:learning_curvse}), so by thermodynamic integration the learnt density matrix has a lower free energy. 
This is the finite temperature analogue of ground-state variational theory, and the model selection criterion advocated by Militzer\cite[\S4.3]{militzer2000THESIS}. 

We were delighted to find that evaluating the restricted PIMC with our fully trained $\rho_\theta$ at $\beta=1.0$ ($1\times10^5$ Monte Carlo steps) produced a thermodynamic energy of $E_{\mathrm{thermo}} \approx 8.715$ and a virial energy of $E_{\mathrm{virial}} \approx 8.490$. 
The thermodynamic estimator is within error of the reference ($E \approx 8.719(3)$), and the convergence of the two estimators seems a further metric of the quality of the nodal structure. 
Given the lack of a nodal action (this and perhaps other missing terms continue to cause errors in the virial estimator), it seems astonishing that a sufficiently accurate learnt density matrix permits the simple thermodynamic estimator to provide an accurate answer. 

Unfortunately our run finished just when the physics was getting interesting and quantum degeneracy was starting to dominate. 
Straightforward efficiency improvements in our codes should shortly enable us to sample a much larger parameter space, and to look at more challenging sign regimes (e.g. $N \gg 3$, high densities).


\section{Discussion}

This method can be considered a modern variational density matrix (VDM)
approach\cite{militzer2000THESIS,MilitzerCeperleyPRE2001,MilitzerPollockPRE2000,MilitzerCeperleyPRL2000},
which similarly minimises the integrated Bloch residual to evolve
a single-Gaussian orbital ansatz from the free-particle limit. 
Here we have a more expressive model for the density matrix (a continuous
normalising flow of quasi-particle backflow).  
By constructing the backflow from equivariant DeepSets\cite{zaheer2017deep}
we retain full many-body Fermionic antisymmetry (VDM truncates to
pair-exchanges only).
By constructing the Jastrow factor from invariant DeepSets
we retain exact $\mathbf{R} \leftrightarrow \mathbf{R}'$ spatial symmetry
(VDM loses this symmetry by enforcing an origin for the ODE). 

Xie et al.\cite{Xie2024} introduce an equivariant normalising flow approach,
targeted at minimising the \emph{free} energy, represented spectrally in an
eigenenergy basis, which renders calculation of the entropy $\rho \ln \rho$
tractable. 
Rather than try and directly learn the observable, here we intentionally
restrict ourselves to learning the density matrix purely to accelerate
a \emph{restricted} path integral Monte Carlo calculation.

However, our method is currently very much proof-of-principle. 

We do not consider the nodal action, so are missing this energy contribution
from our \emph{restricted} PIMC estimators. 
Methods have been developed\cite[Section 2.6.7]{militzer2000THESIS} from finite-differences which offer both distance-based and improved second-order corrections for this energy contribution. 
Our model density matrix has analytic gradients, $\nabla_{\mathbf{R}} \rho_\theta$, which should permit relatively simple implementation of these, and higher-order, techniques. 
These spatial gradients are already calculated to minimise the Bloch residual. 

Because we do not include this action, and perhaps due to
other bugs and errors in the code, we do not reproduce
literature results, or our own canonical PIMC data. 

Following Militzer\cite{militzer2000THESIS} we therefore looked for evidence that
our backflow transform has lowered the \emph{restricted} energy relative to the
\emph{free particle} reference at all temperatures, and found that it made a reasonable improvement. 

\subsection{The homogeneous electron gas}

While the harmonic trap provides proof-of-principle with a simple Hamiltonian and essential closed-form solutions for the non-interacting case, much richer physics could be accessed if we could study the homogeneous electron gas (HEG/UEG) under periodic boundary conditions. 
However, the periodicity introduces significant additional complexities. 

First, the neural velocity field $\bm{v}_\theta$ must be strictly periodic; therefore some kind of periodic feature embedding such as $\sin(2\pi \Delta \mathbf{r}_{ij} / L)$ is required so that the flow is continuous across the boundaries. 
Xie et al.\cite{Xie2024} had some success using such periodic features to guarantee continuity across the cell boundaries.  

Second, the free particle reference ($\beta \to 0$) is no longer a simple real-space Gaussian Mehler kernel, but a periodic kernel on a torus. 
As this decays exponentially on the distance scale of the thermal wavelength, we should be able to evaluate it to sufficient precision with a few nearest-neighbour numerical image cells.

Third, the macroscopic winding numbers accrued by periodic paths would need to be considered in the context of the \emph{restricted} Worm algorithm, particularly whether ergodicity across winding sectors is maintained, and the impact on nodal surface trapping. 

\subsection{Directly using $\rho_\theta$}

An alternative perspective is to view the path integral Monte Carlo simulation as a data generator for training a continuous neural surrogate $\rho_\theta(\R,\R';t)$. 
From this standpoint of probabilistic numerics, we can then ask what physical observables can be extracted directly from this continuous representation without requiring further Monte Carlo sampling. 

For instance, evaluating off-diagonal properties such as the momentum
distribution, natural orbitals, and the electronic excitation spectrum
traditionally requires Monte Carlo sampling with open
paths\cite{militzer2000THESIS} at large computational cost, compounded by ill-posed analytic continuation. 
With a converged continuous representation of $\rho_\theta$, these observables
can be directly evaluated, or diagonalised from, the model. 

There is also the intriguing possibility of using this parametrised density matrix as a generator of low-energy effective models. 
Having defined a basis set, $|\phi_i\rangle$, we can then evaluate the matrix elements of the thermal density matrix in this subspace, 
\begin{equation}
\rho_{ij}(t) = \langle \phi_i | e^{-t \mathcal{H}} | \phi_j \rangle = \int \mathrm{d}\mathbf{R} \, \mathrm{d}\mathbf{R}' \phi_i^*(\mathbf{R}) \, \rho_\theta(\mathbf{R} , \mathbf{R}'; t) \, \phi_j(\mathbf{R}') .
\end{equation}
Because our model $\rho_\theta$ is continuous in imaginary time, we can get back to the Hamiltonian directly via the Bloch equation, $\mathcal{H} \rho_\theta = -\frac{\partial \rho_\theta}{\partial t}$. 
Therefore, the matrix elements of the first-principles Hamiltonian projected onto our effective basis set are simply given by the time-derivative of the model density matrix, which is analytic and should be accessible computationally with automatic differentiation, 
\begin{equation}
\mathcal{H}_{ij}^{\text{eff}} = \int \mathrm{d}\mathbf{R} \, \mathrm{d}\mathbf{R}' \, \phi_i^*(\mathbf{R}) \left( -\frac{\partial \rho_\theta}{\partial t} \right) \phi_j(\mathbf{R}') .
\end{equation}
Therefore once we have a fitted $\rho_\theta$, we can read off a PPP-like Hamiltonian with hopping elements $t_{ij}$ and exchange interaction $J$. 
This provides a form of downfolding\cite{Wagner2016} without having to generate explicit wavefunctions or perform empirical parameter fitting.

\section{Conclusion}

We have introduced \textsc{Spindrift}, a self-consistent algorithm that learns the finite-temperature density matrix on-the-fly from restricted path integral Monte Carlo simulations. 
By learning along a continuous projection of imaginary time from the theoretical purity of the infinite-temperature free-particle reference ($t = 0$) toward quantum degeneracy ($t = \beta \to \infty^+$), we track the continuous emergence of nodal complexity. 
This piecemeal growth of correlation, and the highly informative Bloch residual at each curiculum stage, offers a tractable route to learning the complexity of degenerate quantum matter. 

We provide proof-of-principle simulations of an $N=3$ two-dimensional harmonic trap, with Kelbg screened Coulomb interactions. 
Our learnt density matrix systematically lowers the restricted energy relative to the free-particle reference at each temperature and successfully reproduces the benchmark energy at $\beta=1$, leaving no detectable fixed-node error. 
Despite the the lack of a nodal action in our estimators, our thermodynamic estimator converges to the reference value. 

One interesting observation was that the number of failed reconnections in the Worm algorithm from the open-Worm G-sector to the closed-Worm Z-sector can be used directly as a diagnostic of poor nodal structure, as these Worms get trapped due to the mismatch with the trial nodal structure and the true interacting Hamiltonian which the Worms are feeling. 

Our simple Jastrow factor is perhaps limiting, as the Bloch residual cannot descend below $1 \hbar\omega$ while the spatial standard deviation continues to shrink. 
Incorporating the Coulomb cusps that are clearly visible in the Bloch residual is an obvious thing to try. 
Extending to spin unpolarised systems will perhaps require a more careful consideration of nodal topology\cite{Mitas2006ArXiv}.

By following the Bloch equation along imaginary time, the nodal structure can learn itself.

\begin{acknowledgements}

This work grew out of discussions at the April 2026 \textsc{The Sign Problem of Fermions} workshop at ECT*, Villazzano (Trento), Italy.
I am grateful to the organisers for the motivating invitation and to all the attendees for fruitful academic discussions.

Much understanding underlying this work came from discussions with the groups of Matthew Foulkes, Roberto Bondesan, and members of my research group, at our Machine Learning for Quantum Matter (ML4QM) seminars at Imperial. 
Many thanks to David Ceperley for clarifying some aspects of the VDM by email. 

J.M.F.
is supported by a Royal Society University Research Fellowship
(URF-R1-191292).

Julia codes which implement the permutation-family based PIMC method are available on GitHub\cite{Halcyon.jl}.

I gratefully acknowledge the use of the Imperial College Research Computing Service\cite{HPC}.
\end{acknowledgements}

\bibliography{2026-06-Halcyon-Spindrift-RPIMC}
\end{document}